\documentclass[preprint,showpacs,preprintnumbers,amsmath,amssymb]{revtex4}

\usepackage{doublespace}
\usepackage{graphicx}
\usepackage{dcolumn}
\usepackage{bm}

\begin{document}


\title{Phase Transition in Reconstituted Chromatin}

\author{Tonau Nakai}
\author{Kenichi Yoshikawa\footnotemark}%
\affiliation{%
Department of Physics, Graduate School of Science, Kyoto University \& CREST, Kyoto 606-8502, Japan\\
}%

\author{Kohji Hizume}
\author{Shige. H. Yoshimura}
\author{Kunio Takeyasu}
\affiliation{
Graduate School of Biostudies, Kyoto University, Kyoto 606-8502, Japan\\
}%

\date{\today}

\begin{abstract}
By observing reconstituted chromatin by fluorescence microscopy (FM) 
and atomic force microscopy (AFM), we found that the density of nucleosomes 
exhibits a bimodal profile, i.e., there is a large transition between the dense and dispersed 
states in reconstituted chromatin. Based on an analysis of the spatial distribution of  
nucleosome cores, we deduced an effective thermodynamic potential as a function of the 
nucleosome-nucleosome distance. This enabled us to interpret the folding transition of 
chromatin in terms of a first-order phase transition. This mechanism for the condensation 
of chromatin is discussed in terms of its biological significance.
\end{abstract}

\maketitle
Genomic DNA in eukaryotes is compactly folded into chromatin 
through several hierarchical packings \cite{Woodcock}. The fundamental unit 
of such packing, the nucleosome, consists of 146 bp of DNA 
wrapped around a histone octamer (two molecules each of H2A, 
H2B, H3 and H4). The semi-flexible DNA chain wraps around a 
histone core about two turns \cite{Luger}. It is widely expected 
that the manner of packing and the dynamics of nucleosomes are 
associated with gene activities in living cells \nocite{Wolffe} \nocite{Kornberg} \nocite{Croston} \nocite{Narlikar} [3--6]. 
There have been many studies on the static \nocite{Mangenot} \nocite{Sato} \nocite{Zlatanova} \nocite{Brower-Toland} [7--10]
 and dynamic \cite{Sakaue} properties 
of nucleosomes. It has been shown that nucleosomes condense 
under various conditions; for example, under a high salt concentration 
\cite{Widom}. However, the nature of the higher-order structure is not well 
understood and the underlying physics of nucleosome condensation 
have not yet been clarified. It has been suggested, for example, that 
measurement of the actual interaction energy between nucleosomes 
is essential \nocite{Sato} \nocite{Sakaue} \nocite{Thoma} \nocite{Schiessel} [8,11,13,14] 
for obtaining deeper insight into 
chromatin condensation, but there have been no experimental studies 
on the interaction potential.

On the other hand, it has recently been found that linear DNA larger than 
several tens of base pairs exhibits a large discrete coil-globule transition, 
accompanied by a change in density on the order of $10^4$--$10^5$, upon the 
addition of various kinds of condensing agents \nocite{Mel'nikov} \nocite{Bloomfield} [15,16]. 
It has been revealed 
that this transition is a first-order phase transition under the criterion of 
Landau, i.e., an ON/OFF transition from an elongated coil state to a compact 
globule state \cite{Yoshikawa}. Very recently, such an ON/OFF transition of DNA has been 
suggested to play an important role in gene activity \cite{Yoshikawa}.
Discreteness of the coil-globule transition is a general 
characteristic of single semi-flexible polymer chains \cite{Noguchi}.

The purpose of this study was to obtain a deeper understanding of the 
conformational changes of chromatin, which is essential for obtaining 
insight into genetic activity including duplication, transcription, etc. We 
investigated the physical properties of reconstituted chromatin 
using fluorescence microscopy (FM) and atomic force microscopy (AFM). 
We obtained the pair interaction potential between nucleosome cores from 
an analysis of AFM images and used it to dissect the mechanism of chromatin 
compaction, or chromatin condensation \cite{note}.

The preparation of core histones and 106-kbp plasmids (circular DNA) and the reconstitution of 
chromatin were carried out as previously reported \cite{Hizume}. In this study, the 
mass ratio [histone]/[DNA] was set to 1.0 and 1.3. 
Concentration of NaCl, which affects the interaction between nucleosomes \cite{Mangenot}, 
was 50 mM throughout this article.
Reconstituted chromatin 
samples were fixed with 0.3\% glutaraldehyde in 10mM Hepes-NaOH [pH 7.5] for 30 minutes at 25\symbol{23}C.
The chromatin was placed on a freshly cleaved thin piece of mica 
(thickness; ca. 30--50 $\mu$m) stuck to a glass cover plate (Matsunami 
Glass, No. 1, Japan) for FM observation. The mica surface was pre-treated 
with 10 mM spermidine. This treatment was performed so that chromatin 
would adsorb onto the mica surface. The chromatin complexes were 
visualized by fluorescent microscopy using a fluorescent dye, 0.1 $\mu$M 
4', 6-diamidino-2-phenylindole (DAPI). The observation was performed on a 
droplet (10 $\mu$L) instilled on mica. The sample droplet on mica was 
washed with Millipore water and blown dry with nitrogen gas for 5 minutes. 
Fluorescent chromatin complexes were observed under a Zeiss Axiovert 200 
microscope with a 100$\times$ oil-immersed objective lens at 25\symbol{23}C, and recorded 
on Axio Vision with an AxioCam camera. To obtain two-dimensional real-time 
fluorescent image data, an inverted microscope (IX-70, Olympus) with a 100$\times$ 
oil-immersed objective lens and a highly sensitive EB-CCD camera with an 
image-processing system (Hamamatsu Photonics) were used. The video data 
were recorded on videotapes, and then analyzed with personal computers. 
Due to the blurring effect \cite{Yoshikawa2} in the observation with a 
highly sensitive video system, the size of observed DNA images was assumed 
to be slightly larger (ca. 0.3 $\mu$m) than the actual size of the chromatin. 
The reconstituted chromatin structures observed by FM observation 
were analyzed under Tapping Mode\raisebox{0.5ex}{\scriptsize TM} in AFM (Nanoscope Bioscope, 
Digital Instruments) in air at room temperature. Both FM and AFM 
images were obtained on exactly the same chromatin. The distance 
distribution of nucleosomes was obtained on AFM images acquired with a 
high-resolution AFM apparatus (Nanoscope IIIa).

Figure 1 shows quasi-three-dimensional (3D) FM images (A--C), corresponding 2D images (insets), and 
AFM images (D--F) of reconstituted chromatin adsorbed onto a mica surface. 
The pictures in Fig. 1 (A, D), (B, E), and (C, F) show exactly the same molecules \cite{note2}. The 
mass ratio [histone]/[DNA] is 1.0 in (A, D) and 1.3 in (B, C, E, and F). In (D), 
nucleosomes are dispersed in the chromatin, whereas in (E) 
the condensed and dispersed parts coexist. In (F) the chromatin is 
entirely condensed. Under the condition [histone]/[DNA] = 1.3, the partially 
and entirely condensed states exist in almost equal proportions, while there 
is a very low proportion ($<$ 5\%) of the dispersed state. Under [histone]/[DNA] = 1.0, 
more than 95\% of chromatin is in the dispersed state, while 
the remainder is in the partially condensed state. Despite
 the low resolution, FM observation provides information 
on the degree of condensation of individual reconstituted chromatin in the bulk solution as well as on the surface.

To evaluate the actual size of reconstituted chromatin in bulk solution, 
we measured the Brownian motion of individual chromatin complexes using FM. 
From the time-dependence of the mean square displacement of the center of 
mass of chromatin, we obtained the diffusion constant $D$ using the following relationship:
$\langle (\mbox{\textbf{r}}(0)-\mbox{\textbf{r}}(t))^2 \rangle = 4Dt $ \cite{Matsumoto}.
The hydrodynamic radius $R_H$ is thus deduced as in Table I using the 
Stokes-Einstein relationship, $R_H = k_BT/(6\pi \eta D)$, where $k_B$ is the Boltzmann constant, 
$T$ is the absolute temperature and $\eta $ is the viscosity of the solvent 
(0.89 mPas for pure water at $T$ = 297 K). We also evaluated the size 
of reconstituted chromatin by AFM measurement. The major axis $R_L$ 
and the minor axis $R_S$ of reconstituted chromatin were measured 
in two-dimensional AFM images. Table I shows $R_L$ and $R_S$, together 
with $R_{\mbox{\scriptsize{AFM}}} = \sqrt{R_LR_S}$. 
The relative ratio of $R_{\mbox{\scriptsize{AFM}}}/R_H$ is almost 
the same for [histone]/[DNA] ratios both of 1.0 and 1.3, 
regardless of the difference in the observed state, i.e., the former is 
chromatin on a surface and the latter is chromatin in bulk solution.

We obtained the distribution of distances $N(r)$ as a function of the distance 
$r$ between pairs of nucleosomes in AFM images. It is expected that 
nucleosomes exhibit a nearly equilibrium structure on a 2D mica surface \cite{Yoshinaga}. 
We assumed that the radial distribution function $g(r)$ is proportional to $N(r)/r$. Since the nucleosome density is 
low enough with [histone]/[DNA] = 1.0 (see Fig. 1(A)), the pair potential can be roughly 
deduced from $g(r)$ assuming a Boltzmann distribution, $U(r) = -k_BT\ln g(r)$ 
\nocite{Hansen} \nocite{Han} \nocite{Carbajal-Tinoco} [25--27], where $g(r)$ and $U(r)$ 
were calculated after fitting $N(r)$ to a polynomial. The analysis was performed for 
[histone]/[DNA] = 1.0, since in this condition we can count all of the nucleosomes 
in reconstituted chromatin. The result is shown in Fig. 2 and indicates that $U(r)$ 
has a minimum 0.3 $k_BT$ depth at $r \approx$ 13 nm. By fitting the experimental data in Fig. 2 
with the following equation of potential, $U(r) = -Ar^{-\mu} + Br^{-\nu}$ ($\mu$, $\nu$: integer; $\mu < \nu$), 
we obtain $A$ = 2.4, $B$ = 12, $\mu$ = 9, and $\nu$ = 23, where $r$ is normalized 
with the diameter of a nucleosome (11 nm). These values 
are much larger than those of the Lennard-Jones potential ($\mu$ = 6 and $\nu$ = 12). 
The large values of the exponents can be attributed to the large 
excluded volume of the nucleosome core. A similar profile of potential, i.e., relatively 
large values for the exponents $\mu$ and $\nu$, can also be deduced from an 
analysis of the interaction of neighboring nucleosomes along the DNA chain (data not shown).
Such large values of the exponents can explain the discrete nature of the transition as described below.

Next, we will discuss the conformational stability of chromatin based on 
the above pair interaction energy. The total free energy $F$ of reconstituted 
chromatin with $n$ nucleosomes can be described as
\begin{equation}
F(n) = F_{\mbox{\scriptsize{ela}}} + F_{\mbox{\scriptsize{int}}}(n),
\end{equation}
where $F_{\mbox{\scriptsize{ela}}}$ is the entropic elasticity of the DNA chain and 
$F_{\mbox{\scriptsize{int}}}(n)$ is the volume interaction between nucleosomes. 
We neglect the volume interaction of double-stranded DNA, since the thickness 
of DNA ($\sim$ 2 nm) is much smaller than the diameter of a nucleosome ($\sim$ 11 nm). 
Using the swelling parameter $\alpha $, we obtain 
\begin{equation}
F_{\mbox{\scriptsize{ela}}}/k_BT = \alpha^2 + \alpha^{-2},
\end{equation}
where $\alpha^2 = \langle R^2\rangle /\langle R_0^2\rangle $. $\langle R^2 \rangle$ is 
the mean square of the radius of gyration and $R_0$ is the 
analogous size of an ideal Gaussian coil. $\alpha^2$ and $\alpha^{-2}$ correspond 
to extension and compression of the chain, respectively \cite{Grosberg}. Assuming that $U(r)$ 
has a narrow minimum so that nucleosome-nucleosome interaction occurs only 
among the nearest neighbors, we can take $F_{\mbox{\scriptsize{int}}}(n) \sim nU(r)$. 
By adapting $U(r) = -Ar^{-9} + Br^{-23}$, 
$F_{\mbox{\scriptsize{int}}}(n)$ can be written as
\begin{equation}
F_{\mbox{\scriptsize{int}}}(n, \rho)/k_BT \sim -An\rho ^3 + Bn\rho ^{7.7},
\end{equation}
where $\rho$ is the normalized density of nucleosomes, $\rho \approx r^{-3}$. Thus, we obtain
\begin{equation}
F(n, \rho)/k_BT \sim \alpha^2 + \alpha^{-2} - An\rho^3 + Bn\rho^{7.7}.
\end{equation}
In the present model, $R_0$ decreases with an increase in $n$. 
Thus, we define $L$ as the apparent contour length of chromatin, 
$L = L_0 - an$, where $L_0$ is the contour length of DNA without any histones (36 $\mu$m 
for this sample), and $a$ is the length of DNA wrapped around a histone 
octamer, ca. 50 nm (146 bp). With the Kuhn length $\lambda $ 
(100 nm for a DNA chain) and the number of Kuhn segments $N_S$, 
the size of an ideal Gaussian chain is described as 
$\langle R_0^2 \rangle^{1/2} = \lambda N_S^{1/2}$. Since $N_S = L/\lambda$, we obtain a 
modified $R_0$, which we call $R_0'$, $\langle R_0'^2 \rangle^{1/2} = \lambda (L/\lambda)^{1/2} = \lambda^{1/2}L^{1/2}$. 
Using the relation $\rho \sim nR^{-3}$, we obtain
\begin{equation}
\frac{F(n, \rho)}{nk_BT} = \frac{\rho^{-2/3}}{n^{1/3}\lambda (L_0 - an)} + \frac{\lambda (L_0 - an)}{n^{5/3}}\rho^{2/3} - A\rho^3 + B\rho^{7.7}.
\end{equation}
It is obvious that $F(n, \rho)/nk_BT$ has two minima. The one at the lower-density 
region is derived from the first and second terms in Eq. (5) and the other at 
the higher-density region is derived from the third and fourth terms. As $n$ 
increases, the minimum of the condensed state becomes deeper while a 
double-minimum profile is maintained. When the two minima have a similar 
depth, two different states of high and low nucleosome density coexist. 
Figure 3 shows the free-energy profiles of nucleosomes with $n$ = 400, 500 and 600 
calculated with Eq. (5) together with schemes of the corresponding conformations in 
three dimensions. The observed result regarding the elongated conformation in Fig. 1 
corresponds to the deeper minimum at $\rho \approx 0$. With an increase in $n$, the free energy of 
the condensed state becomes the absolute minimum, which reproduces the experimental 
trend. In the actual experiment for [histone]/[DNA] = 1.3, there are two different states: 
fully condensed and partially condensed. Any instability due to the interfacial energy 
between the condensed and dispersed parts should be negligible considering the 
zero-dimensional nature of the interface. Therefore, we can expect the appearance 
of an intrachain phase-segregated state as in Fig. 1(B). Previous studies on native 
linear DNA chains have indicated that a phase-segregated state is actually observed 
in individual DNA molecules and that the characteristic scale of segregation depends 
on the degree of the surviving electronic charge in the condensed part \nocite{Takagi} \nocite{Ueda} [29,30].

Several biological studies have demonstrated that the higher-order folding of 
chromatin fiber and its dynamic structural changes largely depend on the proper 
functions of various structural and regulatory proteins in the nucleus \nocite{Maeshima} \nocite{Hirano}[31,32], 
which are critical for gene expression and chromosome segregation. On the other 
hand, the results obtained in this study, together with those in several previous 
studies  \nocite{Sakaue} \nocite{Hizume2}[11,33], suggest that the physical properties of a DNA strand 
(length and superhelicity) and the interaction between nucleosomes play 
fundamental roles in chromatin dynamics. 
The higher-order architecture 
of chromatin is determined by the fundamental properties of chromatin fiber 
itself. In this sense, it should be noted that chromosomes are composed 
of several chromatin loops on the order of $\sim$100 kb \nocite{Benyajati} \nocite{Jackson} \nocite{Cook} [34--36], which 
is approximately the same length as used in this study. A first-order 
large-scale conformational transition may explain why previous experiments have 
failed to observe an intermediate state in chromatin condensation. 
More importantly, it may explain the switching of a large number of genes. 
Recently, it has been reported 
that transcription is completely inhibited through an all-or-none 
transition in the structure of giant DNA molecules \nocite{Yoshikawa} \nocite{Tsumoto} [17,37].
Thus, we would like to propose a hypothesis 
of gene self-regulation through a combination of large-scale ON/OFF switching 
due to a conformational transition and several regulatory factors to bind specific regions. 

This work was supported by a Grant-in-Aid for the 21st Century COE ``Center for Diversity 
and Universality in Physics'', the Special Co-ordination Funds (to K. T.), and the COE Research 
Grant from Ministry of Education, Culture, Sports, Science, and Technology of Japan (to K. T.). 
S. H. Y was the recipient of a predoctoral fellowship from the Japan Society for the Promotion of Science.


\newpage

\begin{table}
\caption{\label{tab:table1}The major axis of a chromatin $R_L$, the minor axis $R_S$, $R_{\mbox{\scriptsize{AFM}}}$, 
and the hydrodynamic radius $R_H$. $R_L$ and $R_S$ were measured in AFM images by assuming that 
a chromatin complex is elliptical. $R_H$ was measured in fluorescent images. 
$R_{\mbox{\scriptsize{AFM}}}$ ( = $\sqrt{R_LR_S}$ ) and 
$R_H$ are the hypothetical radii of chromatin if they are considered to be spheres 
on a mica surface and in bulk solution, respectively.}

\begin{ruledtabular}
\begin{tabular}{ccccc}
 [histone]/[DNA] & $R_L$ (nm) & $R_S$ (nm) & $R_{\mbox{\scriptsize{AFM}}}$ (nm) &
 $R_H$ (nm) \\

\hline
1.0 & $420\pm180$ & $250\pm90$ & $320\pm120$ & $370\pm70$\\
1.3 & $220\pm100$ & $130\pm90$ & $170\pm100$ & $190\pm50$\\

\end{tabular}
\end{ruledtabular}
\end{table}

\clearpage

\textbf{Figure Captions}\vspace{1cm}\\
FIG. 1: Reconstituted chromatin with [histone]/[DNA] = 1 
(A, D) and 1.3 (B, C, E, and F). The images in (A, D), (B, E), and (C, F) are photographs of the same chromatin complex. 
(A--C) Light-intensity distribution of fluorescent images of reconstituted chromatin situated 
on a mica surface. Insets are corresponding 2D fluorescent images. 
(D--F) AFM images of the same chromatin as in the fluorescent images. The scale 
bar is 0.2 $\mu$m in AFM images.
\vspace{1cm}\\
FIG. 2: The pair potential of nucleosomes obtained by an equation 
for the Boltzmann distribution. Inset: The nucleosome-nucleosome distance distribution $N(r)$ 
of chromatin reconstituted with [histone]/[DNA] = 1.0. $U(r)$ is deduced by polynomial curve-fitting of $N(r)$.
\vspace{1cm}\\
FIG. 3: Free-energy profiles of a nucleosome with $n$ = 400, 500, and 600 
and corresponding schematic representations, where $n$ is the number of nucleosomes in a 
single chromatin complex. The free energy as a function of the normalized density of nucleosomes 
in the three-dimensional conformation was calculated from Eq. (5).


\begin{thebibliography}{99}
 \bibitem{Woodcock} C. L. Woodcock and S. Dimitrov, Curr. Opin. Genet. Dev. \textbf{11}, 130 (2001).
 \bibitem{Luger} K. Luger, A. W. M\"ader, R. K. Richmond, D. F. Sargent, and T. J. Richmond, Nature \textbf{389}, 251 (1997).
 \bibitem{Wolffe} A. Wolffe, \textit{Chromatin Structure and Function}, 3rd ed. (Academic, London, 1998).
 \bibitem{Kornberg} R. D. Kornberg and Y. Lorch, Cell \textbf{98} 285 (1999). 
 \bibitem{Croston} G. E. Croston and J. T. Kadonaga, Curr. Opin. Cell Biol. \textbf{5}, 417 (1993).
 \bibitem{Narlikar} G. J. Narlikar, H. Y. Fan, and R. E. Kingston, Cell \textbf{108}, 475 (2002).
 \bibitem{Mangenot} S. Mangenot, E. Raspaud, C. Tribet, L. Belloni, and F. Livolant, Eur. Phys. J. E, \textbf{7}, 221 (2002).
 \bibitem{Sato} M. H. Sato, K. Ura, K. I. Hohmura, F. Tokumasu, S. H. Yoshimura, F. Hanaoka, and K. Takeyasu, 
FEBS Lett. \textbf{452}, 267 (1999).
 \bibitem{Zlatanova} J. Zlatanova and S. H. Leuba, J. Mol. Biol. \textbf{331}, 1 (2003).
 \bibitem{Brower-Toland} B. D. Brower-Toland, C. L. Smith, R. C. Yeh, J. T. Lis, 
C. L. Peterson, and M. D. Wang, Proc. Natl. Acad. Sci. USA \textbf{99}, 1960 (2002).
 \bibitem{Sakaue} T. Sakaue, K. Yoshikawa, S. H. Yoshimura, and K. Takeyasu, Phys. Rev. Lett. \textbf{87}, 078105 (2001).
 \bibitem{Widom} J. Widom, J. Mol. Biol. \textbf{190}, 411 (1986).
 \bibitem{Thoma} F. Thoma, T. H. Koller, and A. Klug, J. Cell Biol. \textbf{83}, 403 (1979).
 \bibitem{Schiessel} H. Schiessel, J. Phys.: Condens. Matter \textbf{15}, 699 (2003).
 \bibitem{Mel'nikov} S. M. Mel'nikov, V. G. Sergeyev, and K. Yoshikawa, J. Am. Chem. Soc. \textbf{117}, 2401 (1995). 
 \bibitem{Bloomfield} V. A. Bloomfield, Curr. Opin. Struct. Biol. \textbf{6}, 334 (1996).
 \bibitem{Yoshikawa} K. Yoshikawa, J. Biol. Phys. \textbf{28}, 701 (2002)
 \bibitem{Noguchi} H. Noguchi and K. Yoshikawa, J. Chem.Phys. \textbf{109}, 5070 (1998).
 \bibitem{note} In this paper, we use the term `condensation' to describe the apparent 
aggregation of nucleosomes \textit{in vitro}. In general, the term `aggregation' means random 
concentration, while `condensation' or `compaction' has biological significance and 
reflects ordered packing such as in chromatin condensation. The structures we 
observed are likely to be biologically significant because they mimic the situation \textit{in vivo}. 
Therefore, we would like to use the term `condensation'.
 \bibitem{Hizume} K. Hizume, S. H. Yoshimura, H. Maruyama, J. Kim, H. Wada, and K. Takeyasu, Arch. Histol. Cytol. \textbf{65}, 405 (2002).
 \bibitem{Yoshikawa2} K. Yoshikawa, M. Takahashi, V. V. Vasilevskaya, and A. R. Khokhlov, Phys. Rev. Lett. \textbf{76}, 3029 (1996).
 \bibitem{note2} We performed AFM measurements on reconstituted chromatin with and without DAPI, 
and confirmed that there is essentially no difference between these conditions.
 \bibitem{Matsumoto} M. Matsumoto, T. Sakaguchi, H. Kimura, M. Doi, K. Minagawa, Y. Matsuzawa, and K. Yoshikawa, 
J. Polym. Sci. B: Polym. Phys. \textbf{30}, 779 (1992).
 \bibitem{Yoshinaga} N. Yoshinaga, K. Yoshikawa, and S. Kidoaki, J. Chem. Phys. \textbf{116}, 9926 (2002).
 \bibitem{Hansen} J. P. Hansen and I. R. McDonald, \textit{Theory of Simple Liquids}, 2nd ed. (Academic, San Diego, 1986).
 \bibitem{Han} Y. Han and D. G. Grier, Phys. Rev. Lett. \textbf{91}, 038302 (2003).
 \bibitem{Carbajal-Tinoco} M. D. Carbajal-Tinoco, F. Castro-Rom\'an, and J. L. Arauz-Lara, Phys. Rev. E \textbf{53}, 3745 (1996).
 \bibitem{Grosberg} A. Grosberg and A. Khokhlov, \textit{Statistical Physics of Macromolecules} (American Institute of Physics, NY, 1994).
 \bibitem{Takagi} S. Takagi, K. Tsumoto, and K. Yoshikawa, J. Chem. Phys. \textbf{114}, 6942 (2001).
 \bibitem{Ueda} M. Ueda and K. Yoshikawa, Phys. Rev. Lett. \textbf{77}, 2133 (1996).
 \bibitem{Maeshima} K. Maeshima and U. K. Laemmli, Developmental Cell \textbf{4}, 467 (2003).
 \bibitem{Hirano} T. Hirano, Curr. Opin. Cell Biol. \textbf{10}, 317 (1998).
 \bibitem{Hizume2} K. Hizume, S. H. Yoshimura, and K. Takeyasu, Cell Biochem. Biophys. \textbf{40}, 249 (2004).
 \bibitem{Tsumoto} K. Tsumoto, F. Luckel, and K. Yoshikawa, Biophys. Chem. \textbf{106}, 23 (2003).
 \bibitem{Benyajati} C. Benyajati and A. Worcel, Cell \textbf{9}, 393 (1976).
 \bibitem{Jackson} D. A. Jackson, P. Dickinson, and P. R. Cook, EMBO J. \textbf{9}, 567 (1990).
 \bibitem{Cook} P. R. Cook and I. A. Brazell, J. Cell Sci. \textbf{19}, 261 (1975).
\end{thebibliography}
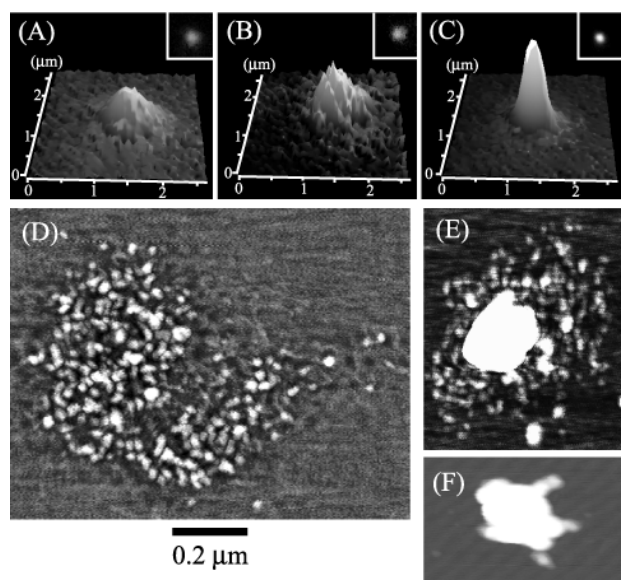
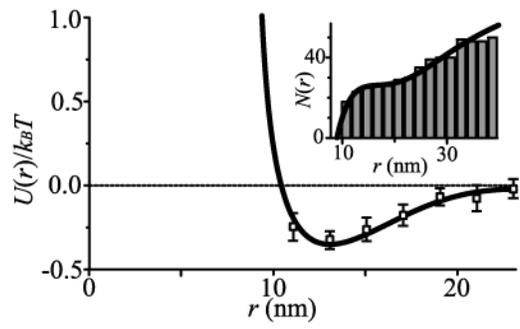
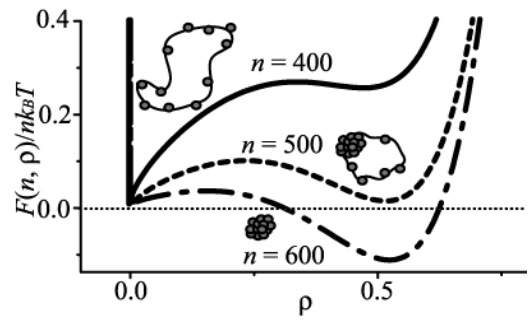
\end{document}